\documentclass[pra,twocolumn,superscriptaddress,showpacs,nobibnotes,notitlepage,nofootinbib]{revtex4-2}

\usepackage[utf8]{inputenc}
\usepackage[margin=1in]{geometry} 
\usepackage{amsmath,amsthm,amssymb,hyperref}
\hypersetup{
    colorlinks=true,
    linkcolor=blue,
    citecolor=magenta,
    filecolor=magenta,      
    urlcolor=blue,
    pdftitle={},
    pdfpagemode=FullScreen,
    }
\usepackage{adjustbox}
\usepackage{graphicx}
\usepackage{float}
\usepackage{xcolor}
\usepackage{soul}
\usepackage{qcircuit}

\usepackage{comment}
\usepackage{booktabs}
\usepackage{multirow}

\usepackage{soul}
\usepackage{xcolor}
\sethlcolor{yellow}

\begin{document}

\title{Experimental Demonstration of High-Fidelity Logical Magic States from Code Switching}

\author{Lucas Daguerre}
\affiliation{Department of Physics and Astronomy, University of California, Davis, CA, 95616, USA}
\author{Robin Blume-Kohout}
\affiliation{Quantum Performance Laboratory, Sandia National Laboratories, Albuquerque, NM 87185, USA}
\author{Natalie C. Brown}
\affiliation{Quantinuum, 303 South Technology Court, Broomfield, Colorado 80021, USA}
\author{David Hayes}
\affiliation{Quantinuum, 303 South Technology Court, Broomfield, Colorado 80021, USA}
\author{Isaac H. Kim}
\affiliation{Department of Computer Science, University of California, Davis, CA, 95616, USA }
\date{\today}

\begin{abstract}
Preparation of high-fidelity logical magic states has remained as a necessary but daunting step towards building a large-scale fault-tolerant quantum computer. One approach is to fault-tolerantly prepare a magic state in one code and then switch to another, a method known as code switching. We experimentally demonstrate this protocol on an ion-trap quantum processor, yielding a logical magic state encoded in an error-correcting code with state-of-the-art logical fidelity. Our experiment is based on the first demonstration of code switching between color codes, from the fifteen-qubit quantum Reed-Muller code to the seven-qubit Steane code. We prepare an encoded magic state in the Steane code with $82.58\%$ probability, with an infidelity of at most $5.1(2.7) \times 10^{-4}$. The reported infidelity is lower than the leading infidelity of the physical operations utilized in the protocol by a factor of at least $2.7$, indicating the quantum processor is below the pseudo-threshold. Furthermore, we create two copies of the magic state in the same quantum processor and perform a logical Bell basis measurement for a sample-efficient certification of the encoded magic state. The high-fidelity magic state can be combined with the already-demonstrated fault-tolerant Clifford gates, state preparation, and measurement of the 2D color code, completing a universal set of fault-tolerant computational primitives with logical error rates equal or better than the physical two-qubit error rate.

\end{abstract}

\maketitle

\section{Introduction}
\label{sec:intro}

Since the discovery of Shor's factoring algorithm~\cite{shor1994algorithms}, it has been well-known that quantum computers can solve certain computational problems exponentially faster than the best known classical algorithms~\cite{shor1994algorithms,lloyd1996universal}. However, the applications for which a quantum computer is expected to provide a substantial advantage tend to require a large number of physical operations~\cite{preskill2018quantum}. Existing quantum computers are noisy, and taming the noise to the level low enough to run these algorithms remains as an outstanding open challenge. 

Theoretically, the threshold theorem asserts that a noisy quantum computer can simulate a noiseless quantum computer with a moderate overhead, if the physical error rate is sufficiently low~\cite{aharonov1997fault}. Bridging the gap between the requirement of the threshold theorem and what is experimentally feasible has been one of the central efforts in quantum computing that spanned multiple decades of work. A large body of recent literature demonstrated different aspects of fault-tolerant quantum computation, e.g., logical state preparation, memory, and gates~\cite{Anderson2021,erhard2021entangling,Postler:2021ddz,bluvstein2024logical,google2023suppressing,reichardt2024demonstration,reichardt2024logical,paetznick2024demonstration}. In particular, recently Ref.~\cite{mayer2024benchmarkinglogicalthreequbitquantum} demonstrated that the fidelity of logical controlled-NOT gates and state preparation are equal or better than their physical counterparts in a trapped ion quantum processor. However there was one logical operation whose fidelity was worse than that of the involved physical operations: $T$-gate, or equivalently, a preparation of logical magic state~\cite{bravyi2005universal}. This is a necessary ingredient for building a universal set of fault-tolerant gates.

Historically, preparation of high-fidelity magic state has been one of the most costly and challenging parts of fault-tolerant quantum computation. A common approach has been to inject a noisy magic state into a code, which is then distilled~\cite{bravyi2005universal}. Injection of a magic state 
is not fault-tolerant, so to reduce the error rate to a level comparable to the other logical operations, it is necessary to carry out magic state distillation~\cite{bravyi2005universal}, which can lead to a considerable amount of extra complexity and overhead; see Ref.~\cite{Litinski2019magicstate} and the references therein. 

An alternative approach is to use code switching~\cite{Anderson2014,bombin2016dimensional,Beverland2021}. In this approach, the magic state is prepared in a code that admits a transversal logical non-Clifford gate, such as the $T$-gate. Then the code is switched to another code that admits a complementary set of fault-tolerant logical gates; see Ref.~\cite{Pogorelov:2024zvv} for a recent experiment. Code switching had been deemed less practical than magic state distillation in the past~\cite{Beverland2021}, but a more recent work~\cite{Daguerre:2024gjd} suggests that it can be improved further than previously thought possible.

In this paper, we experimentally demonstrate that code switching can be used to prepare a high-fidelity magic state encoded in a quantum error-correcting code, using a modest amount of resources. To that end, we present the first experimental demonstration of magic state preparation that employs a code switching between color codes~\cite{Bombin_2dcolorcode,Bombin2007}, from the $[[15,1,3]]$ quantum Reed-Muller code~\cite{steane1996quantumreedmullercodes} to the $[[7,1,3]]$ Steane code~\cite{Steane1996}.  We report the state-of-the-art logical infidelity of at most $5.1(2.7)\times 10^{-4}$, using error-corrected logical quantum state tomography. Indeed, this infidelity is lower than all the reported results in the literature~\cite{Egan:2020kdu,Marques:2021kev,Postler:2021ddz,Anderson2021,Gupta:2023zei,Ye:2023hxg,Kim:2024vmw,Pogorelov:2024zvv,Rodriguez:2024bhh,Lacroix:2024vls}, even the ones whose infidelity is estimated using post-selected logical quantum state tomography~\cite{Lacroix:2024vls}, while only using $28$ qubits in total. In a similar line of research, another experiment done by some of the authors, which uses an error-detecting code and post-selected logical quantum state tomography, will appear elsewhere~\cite{Quantinuum2025}.

Aside from achieving the state-of-the-art infidelity for the logical magic state encoded in a quantum error-correcting code, this experiment also shows that the error rates of the existing quantum processor is below the pseudothreshold of our protocol. The dominant source of errors are the two-qubit error ($1.05(8) \times 10^{-3}$) and the state preparation and measurement (SPAM) error ($1.38(12)\times 10^{-3}$), which are higher than the infidelity of the reported magic state by a factor of $2$ and $2.7$, respectively. Thus, continued improvement in physical layer fidelities should improve the infidelity of the magic state at an even faster rate.

Due to the high fidelity of our magic state, the number of samples needed to certify its fidelity becomes rather large. In order to reduce this complexity, we perform a Bell basis measurement over two copies of the encoded magic state. It was recently shown that such a measurement can lead to an asymptotic improvement in the number of samples needed, assuming the two copies are identical~\cite{Lee:2025lcs}. We use a simple variant of this method with a more relaxed set of assumptions, yielding a rigorous upper bound on the infidelity. A comparison of our results with other experiments in the literature can be found in Table~\ref{tab:comparison}.

\begin{table*}[]
\begin{tabular}{|c|c|c|c|c|c|c|}
\hline
Year                  & Reference                              & State                & Fidelity                       & AR (\%)        & Codes                               & Platform                                                                           \\ \hline
2025                  & \textbf{This work}                     & \textbf{$|\bar{T}\rangle$} & \textbf{$0.99949^{+27}_{-27}$} & \textbf{82.58} & \textbf{Steane, {[}{[}15,1,3{]}{]}} & \textbf{Trapped-ion}                                                                  \\ \hline
\multirow{5}{*}{2024} & \multirow{2}{*}{$^{\text{(PS)}}$Lacroix et al \cite{Lacroix:2024vls}} & $|\bar{T}\rangle$          & $0.9992^{+3}_{-15}$            & 75.2           & \multirow{2}{*}{Steane}             & \multirow{2}{*}{\begin{tabular}[c]{@{}c@{}}Superconducting\\ qubits\end{tabular}}  \\ \cline{3-5}
                      &                                        & $|\bar{H}\rangle $         & $0.9959^{+38}_{-37}$           & 74.6          &                                     &                                                                                    \\ \cline{2-7} 
                      & Rodriguez et al \cite{Rodriguez:2024bhh}                 & $|\bar{M}\rangle$  & $0.994^{+3}_{-4}$              & 1              & Steane, {[}{[}5,1,3{]}{]}           & Neutral atoms                                                                      \\ \cline{2-7} 
                      & $^{\text{(PS)}}$Pogolerov et al \cite{Pogorelov:2024zvv}                 & $|\bar{T}\rangle$          & $0.963^{+4}_{-4}$              & 19             & Steane, {[}{[}10,1,2{]}{]}          & Trapped-ion                                                                          \\ \cline{2-7} 
                      & Kim et al \cite{Kim:2024vmw}                     & $|\bar{H}\rangle$          & $0.8806^{+2}_{-2}$             & 36.3           & Surface code                        & \begin{tabular}[c]{@{}c@{}}Superconducting\\  qubits\end{tabular}                  \\ \hline
\multirow{3}{*}{2023} & \multirow{2}{*}{Ye et al \cite{Ye:2023hxg}}       & $|\bar{T}\rangle$          & $0.8771^{+9}_{-9}$             & 73.41          & \multirow{2}{*}{Surface code}       & \multirow{2}{*}{\begin{tabular}[c]{@{}c@{}}Superconducting\\  qubits\end{tabular}} \\ \cline{3-5}
                      &                                        & $|\bar{H}\rangle$          & $0.9090^{+9}_{-9}$             & 73.41          &                                     &                                                                                    \\ \cline{2-7} 
                      & Gupta el al \cite{Gupta:2023zei}                    & $|\overline{CZ}\rangle$         & $0.9877^{+11}_{-11}$           & 17             & Surface code                        & \begin{tabular}[c]{@{}c@{}}Superconducting\\  qubits\end{tabular}                  \\ \hline
\multirow{3}{*}{2021} & Anderson et al \cite{Anderson2021}                 & $|\bar{T}\rangle$          & $0.978^{+6}_{-6}$              & 100            & Steane                          & Trapped-ion                                                                          \\ \cline{2-7} 
                      & Postler et al \cite{Postler:2021ddz}                  & $|\bar{H}\rangle$          & $0.994^{+5}_{-14}$             & 13.7           & Steane                          & Trapped-ion                                                                           \\ \cline{2-7} 
                      & Marques et al \cite{Marques:2021kev}                 & $|\bar{T}\rangle$          & $0.966$                        & $<$25          & Surface code                        & \begin{tabular}[c]{@{}c@{}}Superconducting\\  qubits\end{tabular}                  \\ \hline
2020                  & Egan et al \cite{Egan:2020kdu}                   & $|\bar{H}\rangle$          & $0.972^{+12}_{-12}$               & 100            & Bacon-Shor                          & Trapped-ion                                                                           \\ \hline
\end{tabular}
\caption{Comparison of experimental magic state fidelities for quantum codes with distance $d\leq 3$, including acceptance rare (AR) and experimental platform. The magic states correspond to: $|\bar{T}\rangle=\frac{1}{\sqrt{2}}(|\bar{0}\rangle +e^{i\pi/4}|\bar{1}\rangle)$, $|\bar{H}\rangle=\cos(\pi/8)|\bar{0}\rangle + \sin(\pi/8)|\bar{1}\rangle$, $|\bar{M}\rangle=\cos(\theta/2)|\bar{0}\rangle + e^{i\pi/4}\cos(\theta/2)|\bar{1}\rangle$ with $\theta=\cos^{-1}(1/\sqrt{3})$, and $|\overline{CZ}\rangle=\frac{1}{\sqrt{3}}(|\overline{00}\rangle+|\overline{10}\rangle+|\overline{01}\rangle)$. Our reported value 0.99949(27) is a lower bound for the fidelity (see Section~\ref{sec:results}). Results marked with (PS) indicate that post-selection is further utilized in the logical state tomography measurements. Using post-selected logical state tomography measurements, our reported fidelity lower bound is 0.99985(11) with 81.26\% of acceptance rate.} \label{tab:comparison}
\end{table*}
\section{Magic state preparation protocol}
\label{sec:protocol}

\begin{figure}[!ht]
    \centering
    \includegraphics[width=0.3\textwidth]{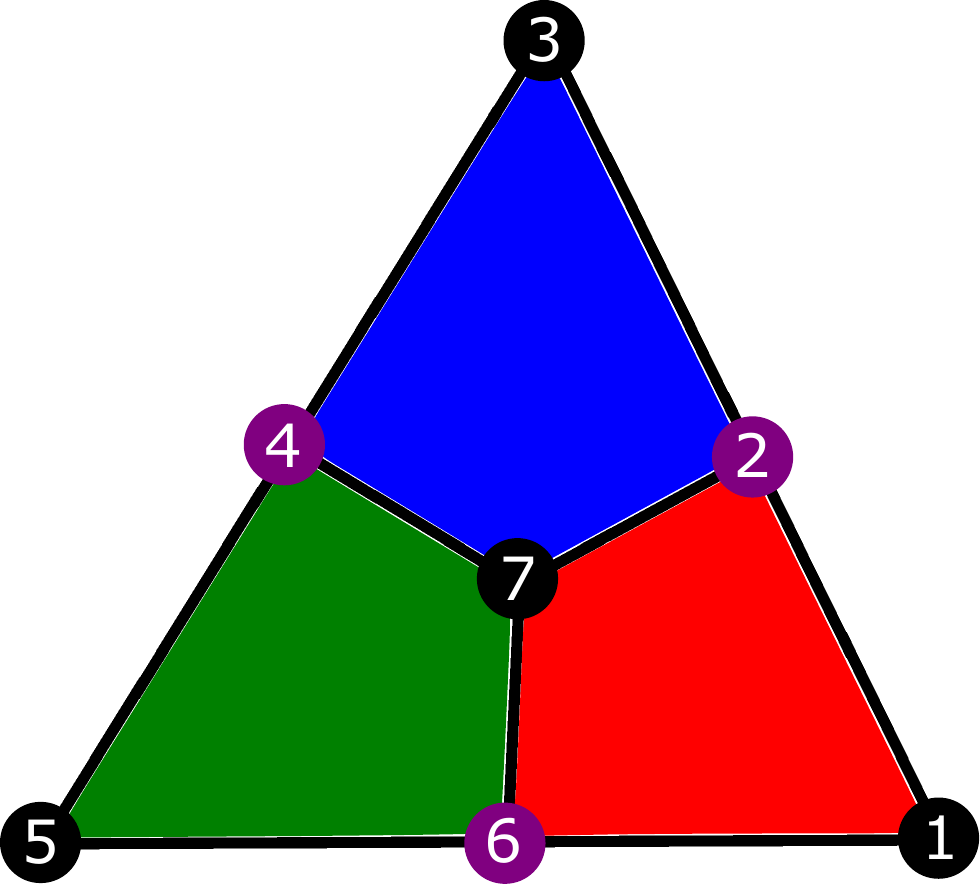}
    \caption{Schematic representation of the $[[7,1,3]]$ Steane code in a 2D lattice including labels for the physical qubits.}
    \label{fig:Steane}
\end{figure}

\begin{figure}[!ht]
    \centering
    \includegraphics[width=0.3\textwidth]{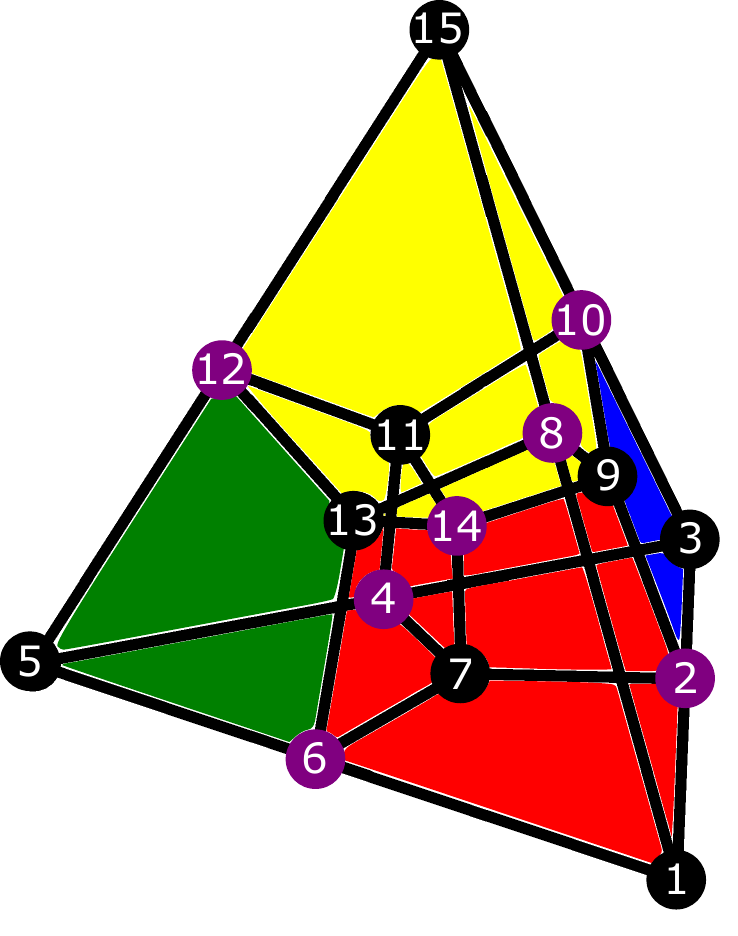}
    \caption{Schematic representation of the $[[15,1,3]]$ qRM code in a 3D tetrahedral lattice including labels for the physical qubits. }
    \label{fig:qRM}
\end{figure}

\begin{figure}[!ht]
\centering
\[
\begin{array}{c}
\centering
\scalebox{1.2}{\quad \quad\quad\quad\quad\quad\Qcircuit @C=1.05em @R=1.0em  {\lstick{\text{qRM:}\:\:\:\:|\bar{+}\rangle} &\gate{\bar{T}}   &\ctrl{1}   & \measuretab{M_{\bar{X}}} & \control \cw &  \\
\lstick{\text{Steane:}\:\:\:\:|\bar{0}\rangle} &\qw  &\targ & \qw & \gate{\bar{Z}} \cwx & \qw & &\lstick{|\bar{T}\rangle}
}}
\end{array}
\]
\caption{The magic state preparation protocol at the logical level~\cite{Daguerre:2024gjd}. A qRM (quantum Reed-Muller) and a Steane codeblock are prepared in $|\bar{+}\rangle$ and $|\bar{0}\rangle$, respectively. The transversal logical $\bar{T}$ gate is applied, creating a logical magic state $|\bar{T}\rangle=\bar{T}|\bar{+}\rangle$ in the qRM code. The magic state is then teleported to the Steane codeblock by applying a logical $\overline{\text{CNOT}}$ gate, measuring the logical $\bar{X}$ operator and applying a logical $\bar{Z}$ correction if $\bar{X}=-1$. The output of the protocol is a logical magic state $|\bar{T}\rangle$ in the Steane codeblock.} \label{fig:logical_circuit} 
\end{figure}
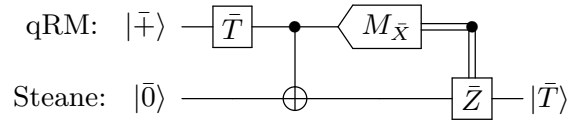

We prepare a magic state $|\bar{T}\rangle=\frac{1}{\sqrt{2}}(|\bar{0}\rangle + e^{i\pi /4} |\bar{1}\rangle)$ using the code switching protocol of Ref.~\cite{Daguerre:2024gjd}. This protocol utilizes two different quantum error-correcting codes: the Steane code [[7,1,3]]~\cite{Steane1996} and the quantum Reed-Muller [[15,1,3]] (qRM) code~\cite{steane1996quantumreedmullercodes}. These codes require $n=7$ and $n=15$ physical qubits, respectively, and encode a single logical qubit $k=1$.  They are also distance $d=3$ instances of the two-dimensional (2D)~\cite{Bombin_2dcolorcode} and three-dimensional (3D)~\cite{Bombin2007} color code families, respectively. Indeed, they are capable of correcting arbitrary single-qubit errors and detecting arbitrary two-qubit errors. While the Steane code supports the full Clifford group transversally, it does not admit any transversal non-Clifford gate~\cite{eastin2009restrictions,bravyi2013}. In contrast, the qRM code supports a transversal non-Clifford $\bar{T}=\text{diag}(1,e^{i \pi/4})$ gate~\cite{Bombin2007}. Because of this, the code-switching protocol of Ref.~\cite{Daguerre:2024gjd}, described at the logical level in Fig.~\ref{fig:logical_circuit}, allows the preparation of a magic state $| \bar{T}\rangle=\bar{T}|\bar{+}\rangle$ in the Steane code. Hence, the magic state can be used to implement a logical $\bar{T}$ gate to an arbitrary logical state encoded in the Steane code via gate teleportation~\cite{Zhou2000}, effectively circumventing the restrictions imposed by~\cite{eastin2009restrictions,bravyi2013}. For completeness, we provide additional information about the color codes in Appendix~\ref{app:color_codes}, as well as a geometrical realization of their embedding in 2D and 3D lattices in Figs.~\ref{fig:Steane} and~\ref{fig:qRM}, respectively.


The protocol of Fig.~\ref{fig:logical_circuit} comprises of several different logical components. Regarding state initialization, a Steane codeblock is prepared in the logical $|\bar{0}\rangle$ state, and a qRM codeblock is prepared in the logical $|\bar{+}\rangle$ state. Their initialization is done using verification flag-based gadgets~\cite{Chao2018,goto2016minimizing,Butt2024,Daguerre:2024gjd}, described in Appendix~\ref{app:color_codes}. Although this process is often viewed as post-selection, we instead refer to this as a \emph{pre-selection}~\cite{paetznick2024demonstration} gadget. Because the magic state preparation protocol can be repeated until success, without affecting the other parts of the computation.

\begin{figure}[!ht]
    \centering
    
\includegraphics[width=0.3\textwidth]{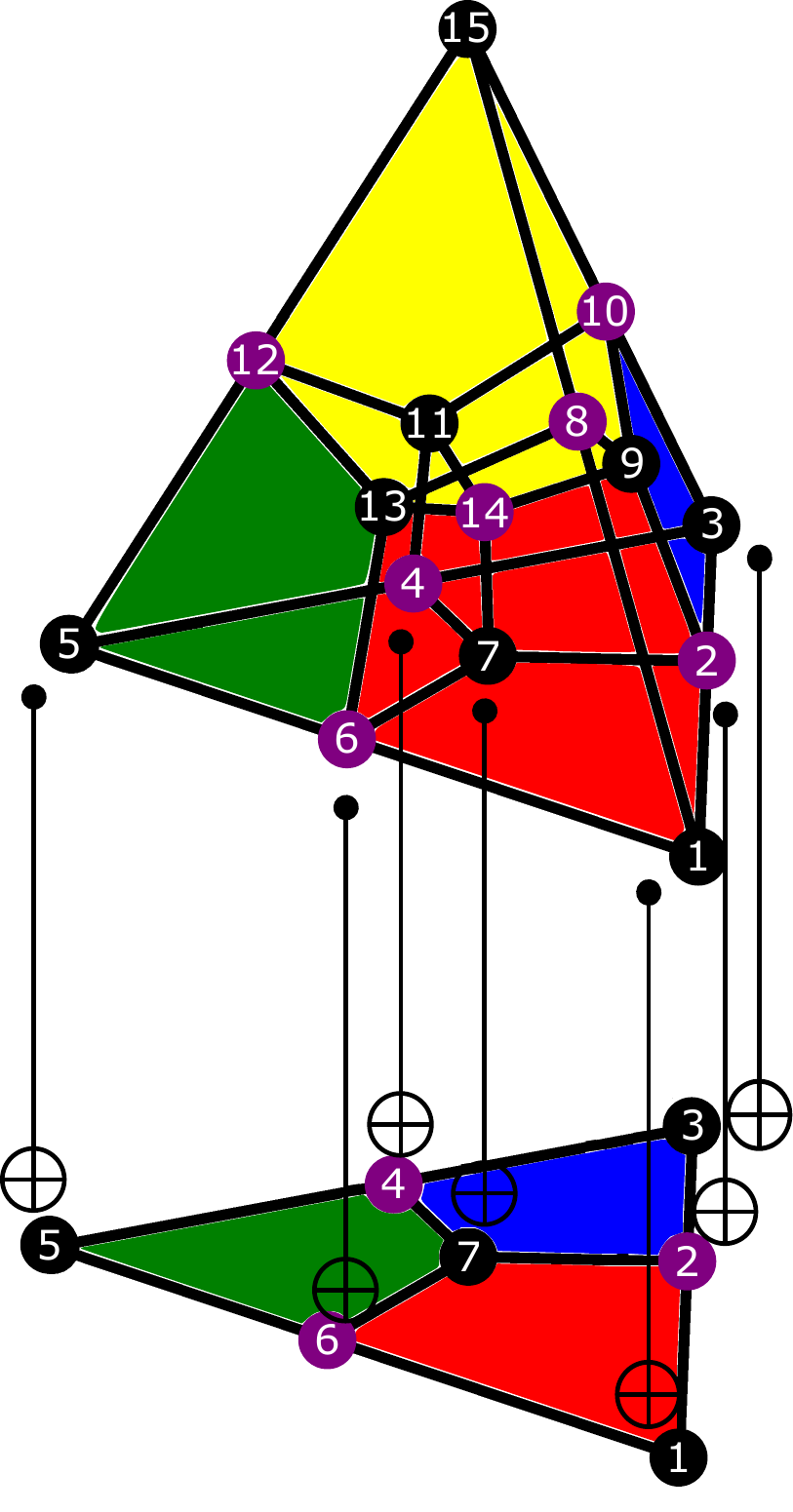}
    \caption{Logical $\overline{\text{CNOT}}$ gate between the Steane and qRM codeblocks, consisting of physical CNOT gates acting on the base of the tetrahedron as controls and on the Steane codeblock as targets.}
    \label{fig:cnot_gate}
\end{figure}

After initializing the two code blocks, a magic state is prepared in the qRM codeblock using the native transversal logical $\bar{T}$ gate given by
\begin{equation}
\bar{T}=T_1T^{\dagger}_2T_3T^{\dagger}_4T_5T^{\dagger}_6T_7T^{\dagger}_8T_9T^{\dagger}_{10}T_{11}T^{\dagger}_{12}T_{13}T^{\dagger}_{14}T_{15}\:.
\label{eq:Tgate}
\end{equation}
Once the magic state is prepared in the qRM codeblock, its logical information is transferred to the Steane codeblock using a logical one-bit teleportation gadget~\cite{Zhou2000}. Correspondingly, an entangling logical $\overline{\text{CNOT}}$ gate between both codeblocks is applied~\cite{sullivan2024,Heuben2024}
\begin{equation}
    \overline{\text{CNOT}} = \bigotimes_{i=1}^7 \text{CNOT}_i\:, \label{eq:logical_cnot}
\end{equation}
which consist of physical CNOT gates between qubits at the bottom face of the qRM codeblock acting as controls, and the Steane codeblock acting as targets, see Fig.~\ref{fig:cnot_gate}. Finally, the logical $\bar{X}$ operator is destructively measured in the qRM codeblock, where further pre-selection is employed (see Appendix~\ref{app:exp_methods}). In cases where the measurement outcome $\bar{X}=-1$, a logical $\bar{Z}$ correction has to be applied, which in practice it can be taken into account in software post-processing without being physically applied (see Appendix~\ref{app:exp_methods}).

\section{Logical magic state certification}
\label{sec:state_tom}
\begin{figure}[!ht]
    \centering    
\includegraphics[width=0.4\textwidth]{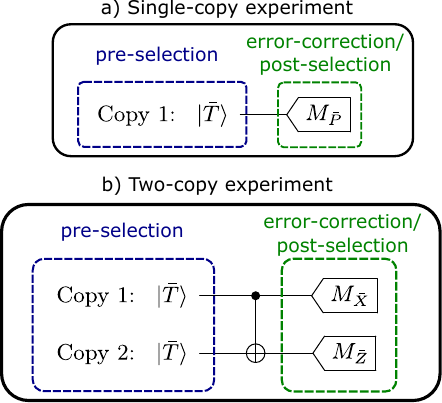}
    \caption{Single-copy (a) and two-copy experiments (b), respectively. In both experiments, the magic states are prepared using pre-selection as in Fig.~\ref{fig:logical_circuit}. In (a) the logical $\bar{P}$ operators with $\bar{P}\in 
    \{\bar{X},\bar{Y},\bar{Z}\}$ are measured, whereas in (b) the logical $\bar{X}_1\otimes \bar{Z}_2$ operator is measured. The overlap with the singlet $\epsilon$ (\ref{eq:epsilon_def}) is given by the fraction of measurement outcomes such that $(m_{\bar{X}},m_{\bar{Z}})=(-1,-1)$. In both cases, the logical operators can be determined either with an ideal round of error-correction or error-detection/post-selection.}
    \label{fig:two_copy_protocol} 
\end{figure}

In order to quantify the quality of the noisy magic state $\bar{\rho}$ produced in the experiments, we perform quantum state tomography at the logical level to estimate the fidelity $F(\bar{\rho},|\bar{T}\rangle \langle \bar{T}|)$. The fidelity can be simplified to the following expression (see (\ref{eq:fid_delta}) in Appendix~\ref{app:tomography}):
\begin{equation}
    F(\bar{\rho},|\bar{T}\rangle \langle \bar{T}|)=\frac{1}{2}+\frac{1}{2\sqrt{2}}\big(\langle \bar{X}\rangle_{\bar{\rho}}+\langle\bar{Y}\rangle_{\bar{\rho}}\big)\:,
\label{eq:fid_formula}
\end{equation}
where  $\langle \bar{P}\rangle_{\bar{\rho}}=\text{Tr}(\bar{\rho}\bar{P})$ is the expected value of $\bar{P}\in \{\bar{X},\bar{Y}\}$. We refer to the method of determining $\langle \bar{P}\rangle_{\bar{\rho}}=\text{Tr}(\bar{\rho}\bar{P})$ for $\bar{P}\in \{\bar{X},\bar{Y},\bar{Z}\}$ as the \emph{single-copy experiment}, see Fig.~\ref{fig:two_copy_protocol}a. These expected values can be estimated by applying physical measurements in the $X$, $Y$, or $Z$ eigenbasis and applying error-correction or error-detection/post-selection to the measurement outcomes. For more details, see Appendix~\ref{app:exp_methods}.

A drawback of determining the fidelity solely from (\ref{eq:fid_formula}) is that neither $\bar{X}$ nor $\bar{Y}$ possess the target magic state as an eigenstate. Thus, their variance is of order unity. If we obtain $N_{\text{shots}}$ independent shots to determine the expected values, the fidelity can be bounded only up to $F\geq 1- O(1/\sqrt{N_{\text{shots}}})$. This is unpractical if the infidelity $1-F$ is small, which is precisely the case at hand.

In order to lower the sample complexity, we employ an alternative method that can achieve $F\geq 1- O(1/N_{\text{shots}})$. This involves performing an additional experiment using two copies of the state, which we refer to as the \emph{two-copy experiment}. The main idea is to perform a simplified version of the SWAP test~\cite{barenco1997stabilization} to estimate the purity of the state; see Fig.~\ref{fig:two_copy_protocol}b for the circuit diagram. This process lets us determine the purity of an experimental state by quantifying the overlap with the singlet (anti-symmetric) state $|\overline{\Phi}_1 \rangle =\frac{1}{\sqrt{2}}(|\overline{01}\rangle-|\overline{10}\rangle)$,
\begin{equation}
\begin{split}
    \epsilon &\equiv \langle \overline{\Phi}_1 | \bar{\rho}_1 \otimes \bar{\rho}_2|\overline{\Phi}_1\rangle\\ &=\langle \overline{-1}|\overline{\text{CNOT}_{s}}(\bar{\rho}_1 \otimes \bar{\rho}_2)\overline{\text{CNOT}_{s}}|\overline{-1}\rangle\:,
\label{eq:epsilon_def}
\end{split}
\end{equation}
where $|\overline{-1}\rangle \equiv |\overline{-}\rangle_1 |\bar{1}\rangle_2$. Here the logical $\overline{\text{CNOT}_{s}}$ gate is implemented between two copies of the Steane code, unlike the logical $\overline{\text{CNOT}}$ (\ref{eq:logical_cnot}) gate; see Appendix~\ref{app:Steane_code} for more details. For the experimental details of the two-copy experiment, see Appendix~\ref{app:exp_methods}.

Combining the results from one- and two-copy experiments, we can obtain a rigorous lower bound on the fidelity:
\begin{equation}
     F(\bar{\rho},\bar{\sigma}_T) \geq  1-\left(\epsilon+\frac{\delta_X^2+\delta_Y^2+\delta_Z^2}{4}\right)\:.
\label{eq:fid_bound}
\end{equation}
The values $\vec{\delta}=(\delta_X,\delta_Y,\delta_Z)$ correspond to the difference of the vectors associated to the Bloch sphere representation of the experimental state $\bar{\rho}=\frac{I+\vec{v}\cdot \vec{\sigma}}{2}$ and the target logical pure state $\bar{\sigma}_T=\frac{I+\vec{u}\cdot \vec{\sigma}}{2}$,
\begin{equation}
    \vec{v}=\vec{u}+\vec{\delta}\:,
\end{equation}
where $\vec{\sigma}$ is the vectorial representation of the Pauli matrices. Note that $\vec{v}=\big(\langle \bar{X}\rangle_{\bar{\rho}},\langle \bar{Y}\rangle_{\bar{\rho}},\langle \bar{Z}\rangle_{\bar{\rho}}\big)$ and for magic $|\bar{T}\rangle \langle \bar{T}|$ states $\vec{u}=\big(\frac{1}{\sqrt{2}},\frac{1}{\sqrt{2}},0\big)$; see Appendix~\ref{app:tomography} for a derivation. The formula  (\ref{eq:fid_bound}) holds regardless of the logical pure target state $\bar{\sigma}_T$, and achieves the desired $F\geq 1- O(1/N_{\text{shots}})$ scaling.

We remark that a recent work also proposed a method with the same scaling in sample complexity~\cite{Lee:2025lcs}. Our approach differs from that work in two aspects. First, unlike Ref.~\cite{Lee:2025lcs}, our result can be extended to the setup in which the states are not necessarily identical, only assuming they are in a product state. In this case, Eq.~\eqref{eq:fid_bound} applies to at least one of the two copies. For the worst-case fidelity between the two copies, we obtain a similar bound, albeit with a worse constant appearing in front of $\epsilon$. Second, Ref.~\cite{Lee:2025lcs} proposes to twirl both copies of the magic states before applying the entangling $\overline{\text{CNOT}_s}$ gate. We forgo the twirling operation and instead perform the single-copy experiments. For future work, it will be interesting to compare the estimated fidelity using these two different approaches.

\section{Experimental results}
\label{sec:results}
The logical state certification experiments were carried out on the Quantinuum H2-1 trapped-ion quantum processor~\cite{Wineland98,Moses2023} using $28$ qubits total ($22$ data and $6$ ancilla qubits) for the single-copy experiments, and $56$ qubits total ($44$ data and $12$ ancilla qubits) for the two-copy experiments. The previous numbers assume mid-circuit measurement and qubit reuse since the single-copy experiments require $31$ qubits ($22$ data and $9$ ancilla qubits), whereas the two-copy experiments require $62$ qubits ($44$ data and $18$ ancilla qubits). A virtue of the trapped-ion quantum processor is its all-to-all connectivity, capable of implementing requisite non-local gates with little additional error. Technical information about the Quantinuum H2-1 quantum processor can be found in Appendix~\ref{app:exp_setup}. 

The fidelity lower bound for the magic state $|\bar{T}\rangle$ is computed using (\ref{eq:fid_bound}). The result gives for an error-corrected magic state an infidelity $1-F\leq 5.1(2.7)\times 10^{-4}$. With a total of $N_{\text{shots}}=10,000$ shots for each Pauli basis, the average acceptance rate across al single-copy experiments is $82.58\%$. The dominant contribution to the fidelity bound comes from the two-copy experiments since $\epsilon=4.56 (2.63) \times 10^{-4}$. We also perform $N_{\text{shots}}=10,000$ shots of the two-copy experiments. The acceptance rate is $65.73\%$ and $3$ logical $|\overline{-1}\rangle$ states are detected in the experiments. The magnitude of $\epsilon$, which quantifies the purity of the state, is one order of magnitude larger than the single-copy experiments contribution to (\ref{eq:fid_bound}), which is $\frac{1}{4}(\delta_X^2+\delta_Y^2+\delta_Z^2)=5.24(5.60) \times 10^{-5}$. A summary of the experimental results can be also found in Table~\ref{tab:exp_results} in Appendix~\ref{app:exp_methods}.

If we employ error-detection/post-selection in the decoding step, the infidelity bound becomes $1-F\leq 1.5(1.1)\times 10^{-4}$, with an acceptance rate of $81.26\%$. The two-copy experiment yields $\epsilon=9.28(9.28) \times 10^{-5}$ with an acceptance rate of $61.82\%$ (no logical $|\overline{-1}\rangle$ states are detected). The contribution to the fidelity bound due to $\epsilon$ is still dominant, because it is almost twice as large as the single-copy experiment contribution $\frac{1}{4}(\delta_X^2+\delta_Y^2+\delta_Z^2)=5.76(5.93) \times 10^{-5}$. Thus, the infidelity may be estimated to be even lower if we take more samples. We leave such studies for future work.

The reported infidelity of $5.1(2.7)\times 10^{-4}$ is below the leading error rate of the involved physical operations. For comparison, the two-qubit error rate is $p_2= 1.05(8) \times 10^{-3}$ and the SPAM error rate $1.38(12)\times 10^{-3}$. Thus these error rates are below the pseudo-threshold of our scheme. Moreover, preparation of the encoded magic state and its characterization has an infidelity expected to be lower than the infidelity of the same process at the physical level. (Such a process is essentially limited by the SPAM error of $1.38(12)\times 10^{-3}$.) Lastly, the reported infidelity is the new state-of-the-art, even compared against the other methods in the literature that use post-selection; see Table~\ref{tab:comparison}. Thus our experiment serves as a new frontier for preparing high-fidelity magic states in experiments, an essential ingredient of fault-tolerant quantum computation.

\section{Discussion}
\label{sec:discussion}

In this paper, we experimentally demonstrate one of the key building blocks of fault-tolerant quantum computation: preparation of high-fidelity encoded magic states. We use code switching from the $15$-qubit qRM code to the $7$-qubit Steane code~\cite{Anderson2014,bombin2016dimensional,Beverland2021}, employing a verified pre-selection protocol~\cite{Daguerre:2024gjd}. The reported infidelity of the logical magic state is the new state-of-the-art result for magic states encoded in a quantum error-correcting code. Moreover, this infidelity is lower than the leading physical error rate (dominated by SPAM) by a factor of $2.7$, indicating that the quantum processor is below the pseudo-threshold of the protocol. 

Combined with the fault-tolerant Clifford gates, state preparation and measurement using the same code~\cite{mayer2024benchmarkinglogicalthreequbitquantum}, we complete a universal set of logical operations whose logical error rates are all equal or better than the physical two-qubit error rate. 

It is notable that the state-of-the-art fidelity was achieved using a rather modest number of qubits ($22$ data and $6$ ancilla qubits). The modest resource requirement suggests that the realistic cost of magic-state distillation may be reduced substantially in the near future. For instance, if we use the standard $15$-to-$1$ magic state distillation protocol~\cite{bravyi2005universal}, the error rate of the magic state can be reduced from $p$ to $\sim 35p^3$. Plugging in the infidelity of our magic state, we obtain an expected magic state infidelity of at most $4.6 \times 10^{-9}$. This is a number low enough to study the challenging 2D Hubbard model~\cite{campbell2021early} and it is even lower than the inverse of the number of $T$-gates to run quantum algorithms for FeMoco~\cite{reiher2017elucidating,li2019electronic,low2025fastquantumsimulationelectronic}. We anticipate this finding to affect future studies of resource estimate for quantum algorithms on a realistic quantum computer.

Our work also highlights the importance of verified pre-selection in designing fault-tolerant quantum computation protocols~\cite{goto2016minimizing,Itogawa:2024zov,gidney2024cult,paetznick2024demonstration,Daguerre:2024gjd}. Certain parts of quantum computations can be repeated probabilistically until success, without affecting other parts of the computation. Our work shows that there is a large design space for improving the existing protocols this way, even in the near-term.

One might suggest the next frontier is to demonstrate a logical $\bar{T}$-gate with an infidelity lower than the physical $T$-gate error rate.  Because the physical single-qubit error rate of H2-1 quantum processor is $1.9(4)\times 10^{-5}$, the magic state created in this experiment is not sufficient for this goal. If the logical error rate is dominated by two-qubit gate errors $p$, and we assume the logical error rate is proportional to $\propto p^2$ in the $d=3$ protocol, the two-qubit gate error would need to improve by approximately $\sqrt{\frac{5.1\times 10^{-4} }{1.9\times 10^{-5}}}\sim 5$x for this goal. Improving the two-qubit gate fidelity by $5$x is not inconceivable, as trapped-ions have achieved gate errors near this level using microwave gates~\cite{Loschnauer:2024gat}; however, such gates have yet to be demonstrated in a system of many interacting qubits.

Extending our approach to a higher-distance instance of the color code may prove to be the most straightforward approach once more qubits are available in next-generation devices. In such a case, because the number of physical qubits scales as $\Theta(d^3)$, and the post-selection rate is going to be exponentially suppressed (although strategies with a partial level of post-selection could be engineered), we expect the role of our protocol to be slightly different than magic state distillation, since the size of the 3D color code should be capped. Indeed, our protocol can be seen as a source of high-fidelity magic states for 2D color codes at some finite distance, which will need to be further complemented with a procedure that increases the code distance, together with at least one round of magic state distillation.

Furthermore, we advise some healthy skepticism in the definition of the next frontier regarding a logical $\bar{T}$-gate breaking-even the physical $T$-gate error rate, as one can imagine this goal being easier to achieve in a system with a slightly worse single-qubit gate error (since the logical error could still be dominated by the two-qubit gate error), and it seems troublesome to define milestones that are easier to achieve with lower performance machines. Another approach to gauging progress is that once all logical operations have error rates below the dominant source of noise in the physical layer (which we have now achieved), the next frontier is to demonstrate algorithmic break-even through lower circuit infidelities, where the circuits use a universal gate set in a standardized way. Of course, defining such a family of circuits is a complex task and beyond the scope of this work.
\newline
\indent Finally, extending our approach to other code families that admit a transversal logical non-Clifford gate such as $\overline{CCZ}$~\cite{Kubica:2015mta}, could be another interesting direction to pursue in the future.

\section*{Acknowledgments}
We thank Ciar\'{a}n Ryan-Anderson for discussions. This research used resources of the Oak Ridge Leadership Computing Facility at the Oak Ridge National Laboratory, which is supported by the Office of Science of the U.S. Department of Energy under Contract No. DE-AC05-00OR22725.

\section*{Authors contribution}
L.D. and I.K. devised the project and designed the protocol implementation. N.B. and D.H. helped with adapting the protocol to the hardware. L.D. implemented the code for circuit generation. L.D. and N.B. executed experiments on the quantum computer and collected data. L.D. Analyzed the data. R.B.-K. devised the sample-efficient magic state certification protocol. L.D. improved bounds for the magic state fidelity bound. L.D. and I.K. contributed to the writing of the manuscript. I.K. led the overall project as the lead principal investigator. All authors contributed to the technical discussions and editing of the manuscript.

\section*{Competing interests}
The authors declare no other competing interests.

\section*{Data availability}
All the datasets and codes can be made available upon
reasonable request to the corresponding author.

\bibliography{bib_magic_exp}
\bibliographystyle{utphys}

\clearpage
\appendix
\section{Experimental setup}
\label{app:exp_setup}
We used Quantinuum's H2-1 trapped-ion processor based on a QCCD architecture~\cite{Moses2023} to perform the experiments in this article. Regarding the quantum proccessor, the most recent randomized benchmarking experiments averaged over all four gate zones showed fidelities of $p_1=1.9(4)\times 10^{-5}$ and $p_2=1.05(8)\times 10^{-3}$ for single-qubit and two-qubit gates, respectively. SPAM errors are dominated by $|1\rangle$ state measurement errors $p_{|1\rangle}=1.38(12)\times 10^{-3}$, over $|0\rangle$ state measurement errors $p_{|0\rangle}=6.0(8)\times 10^{-4}$. Crosstalk
errors from mid-circuit SPAM operations are $p_{\text{ct}}=6.6(8)\times 10^{-6}$. 

Memory errors per qubit were benchmarked in Ref.~\cite{Moses2023} using random circuits, resulting in an estimate for memory error per circuit layer ($\sim60$ ms) of $p_{\text{idle}}=2.0(2) \times 10^{-4}$. However, the circuits used in this work have a very different structure, with some qubits idling for multiple circuit layers. To mitigate coherent accumulation of memory error, we apply dynamical decoupling pulses approximately every $30$ ms, ideally keeping the memory error in each layer of gates $\leq2.0(2) \times 10^{-4}$ while adding a small amount of error from the decoupling pulses ($\sim1.9(4)\times 10^{-5}$). Confirming these effects is difficult in simulation due to the incompatibility of stabilizer simulations with coherent error accumulation, and a comprehensive experimental characterization of individual error sources' impacts is beyond the scope of this work.

The experiments were prepared using the OPENQASM2.0 programming language, designed to represent quantum circuits. The quantum circuits are then converted to the $\texttt{Circuit}$ class using the $\texttt{pytket}$ module, so they can be submitted to the H-series compiler stack where they are converted into machine level instructions. In this process, no optimization is utilized. The quantum circuits are also designed such that no classically-controlled operations are applied, thus minimizing the latency time, though we note that previous experiments~\cite{Anderson2021} have shown that real-time application of corrections can be done without significant error.

\section{Experimental methods}
\label{app:exp_methods}
In this Appendix we provide additional information about the data analysis. A summary of the experimental results and numerical emulations is shown in Table~\ref{tab:exp_results}.

\begin{table*}[!ht]
\begin{tabular}{|c|cc|cc|}
\hline
                                      & \multicolumn{2}{c|}{Error-correction}                              & \multicolumn{2}{c|}{Post-selection}                            \\ \hline 
Quantity                              & \multicolumn{1}{c|}{Value}                        & Acceptance rate$(\%)$ & \multicolumn{1}{c|}{Value}                     & Acceptance rate$(\%)$ \\ \hline \hline
$F(\bar{\rho},|\bar{T}\rangle \langle \bar{T}|)$                                   & \multicolumn{1}{c|}{$0.99949(27)$}                & $82.58$  & \multicolumn{1}{c|}{$0.99985(11)$}                         &    $81.26$      \\ \hline \hline
$\epsilon$                            & \multicolumn{1}{c|}{$4.56 (2.63) \times 10^{-4}$} & $65.73$  & \multicolumn{1}{c|}{$9.28(9.28) \times 10^{-5}$} & $61.83$  \\ \hline
$\langle \bar{X}\rangle_{\bar{\rho}}$ & \multicolumn{1}{c|}{$0.6993(78)$}                 & $83.56$  & \multicolumn{1}{c|}{$0.6987(78)$}                          &    $82.66$      \\ \hline
$\langle \bar{Y}\rangle_{\bar{\rho}}$ & \multicolumn{1}{c|}{$0.7193(77)$}                 & $81.24$  & \multicolumn{1}{c|}{$0.7197(77)$}                          &     $79.72$     \\ \hline
$\langle \bar{Z}\rangle_{\bar{\rho}}$ & \multicolumn{1}{c|}{$0.0000(109)$}               & $82.96$  & \multicolumn{1}{c|}{$-0.0007(110)$}                          &    $81.42$      \\ \hline
\end{tabular}
\caption{Experimental results for the single-and two copy experiments analyzed using an ideal round of error-correction or post-selection/error-detection. The fidelity lower bound is computed using (\ref{eq:fid_bound}). The fidelity acceptance rate is the average over the acceptance rates of the single-copy experiments.} 
\label{tab:exp_results}
\end{table*}

In the single-copy experiments (Fig.~\ref{fig:two_copy_protocol}) the output of a given shot is a binary string with $31$ entries. Out of those $31$ bits, $7$ of them correspond to Steane code data qubits, $15$ to qRM code data qubits, $4$ to $Z$-type qRM code stabilizers, and $5$ to flags ($2$ from state initialization, and $3$ from the $Z$-type stabilizer circuit measurements). We pre-select and analyze only those shots such that the $4$ $Z$-type stabilizers and $5$ flags are $+1$. Moreover, from the $15$ entries corresponding to the $X$-type destructive measurements of the qRM code data qubits, we extract the qRM $X$-type syndrome using the definitions shown in Eq.~(\ref{eq:3cells}). In case any of these $X$-type syndromes are non-trivial, we discard the shot. Finally, the information from the remaining $7$ Steane code data qubits is analyzed to extract logical Pauli basis expected values $\langle \bar{P}\rangle_{\bar{\rho}}$ for $\bar{P}\in \{\bar{X},\bar{Y},\bar{Z}\}$. First, the Steane code syndrome is extracted using (\ref{eq:steane_plaquettes}), followed by a correction which is applied based on a look-up table decoder (in cases where post-selection is considered, only the instances with trivial syndromes are analyzed). Then, the logical operator $\bar{P}$ of the single-copy experiment is computed using the definitions shown in Eq.~(\ref{eq:Steane_logicals}). Since the logical $\bar{Z}$ correction of Fig.~\ref{fig:logical_circuit} is not physically applied, when the logical $\bar{X}^{\text{qRM}}=-1$ in the qRM code, the resultant logical operator $\bar{P}$ is flipped if $\bar{P}=\bar{X}\text{ or }\bar{Y}$.

Similarly, the data analysis of the two-copy experiment of Fig.~\ref{fig:two_copy_protocol} entails decoding a $62$ binary string for every shot. Only shots where the qRM $X$ and $Z$-type stabilizers and flags are $+1$ for both copies are kept. Thus, the acceptance rate of the two-copy experiment is roughly the square of the acceptance rate of the single-copy experiment. To extract the logical $\bar{X}_1\otimes \bar{Z}_2$ operator, similar to the decoding process of the single-copy experiment, $\bar{X}_1$ and $\bar{Z}_2$ are obtained for the respective Steane code copy. Since the logical $\bar{Z}$ corrections of the two copies of the protocol of Fig.~\ref{fig:logical_circuit} are not physically applied, they are considered in post-processing. Because $\overline{\text{CNOT}_{s}}(\bar{Z}_1^{a_1}\otimes \bar{Z}_2^{a_2})\overline{\text{CNOT}_{s}}=\bar{Z}_1^{a_1+a_2}\otimes \bar{Z}_2^{a_2}$ (where $a_i=(1-\bar{X}_i^{\text{qRM}})/2$ indicates whether a correction has to be applied to the $i$-th copy), the logical measurement of the first copy $\bar{X}_1$ has to be flipped when $a_1+a_2 \equiv 1 \pmod 2$. 

The values reported in Table~\ref{tab:exp_results} correspond to the mean over all the post-selected shots $N_{\text{post}}$, and the acceptance rate is the number of post-selected shots over the total number of shots $N_{\text{shots}}$. The reported acceptance rate for the fidelity in Tables~\ref{tab:comparison}~and~\ref{tab:exp_results} is the average over the acceptance rates of the single-copy experiments.

The (1$\sigma$) confidence error bars are computed as $\Delta \langle \bar{P}\rangle_{\bar{\rho}}=\sqrt{\frac{\text{Var}(\bar{P})}{N_{\text{post}}}}$ for the single-copy experiments, where $\text{Var}(\bar{P})$ is the sample variance. For the two-copy experiment, the (1$\sigma$) error bar is computed using a Binomial distribution $\Delta \epsilon=\sqrt{\frac{p_f (1-p_f)}{N_{\text{post}}}}$, where $p_f=\frac{N_{|\overline{-1}\rangle}}{N_{\text{post}}}$ is the fraction of detected logical $|\overline{-1}\rangle$ states $N_{|\overline{-1}\rangle}$. In case where $N_{|\overline{-1}\rangle}=0$, the ($1\sigma$) confidence interval is $[0,\frac{1.147}{N_{\text{post}}})$ \cite{Hanley1983IfNG}, so we choose $p_f=\Delta \epsilon=\frac{1.147}{2N_{\text{post}}}$. Finally, the $(1\sigma)$ error bar for the fidelity lower bound for $F$ (\ref{eq:fid_bound}) is computed assuming uncorrelated error propagation.

\section{Color codes}
\label{app:color_codes}
In this Appendix we describe properties of the Steane [[7,1,3]] code  (Appendix~\ref{app:Steane_code}) and the quantum Reed-Muller [[15,1,3]] (qRM) code (Appendix~\ref{app:qRM_code}).

\subsection{Steane code}
\label{app:Steane_code}

\begin{figure*}[!ht]
    \centering
    \includegraphics[width=0.8\textwidth]{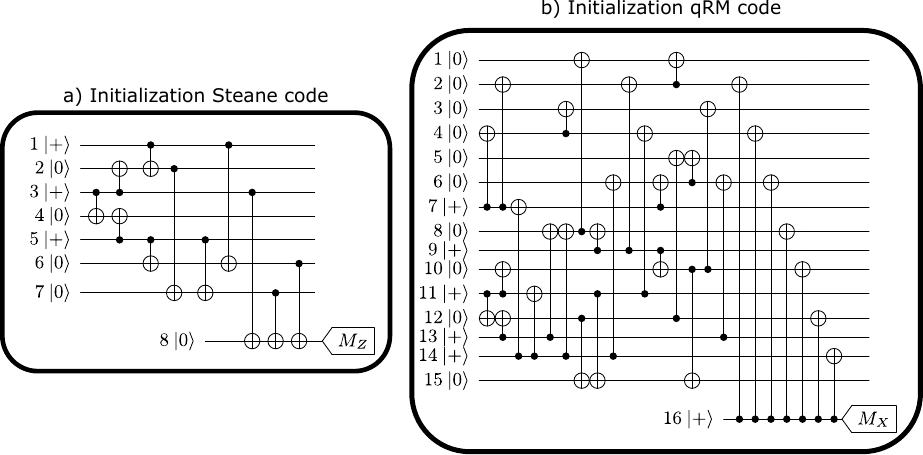}
    \caption{(a) Flag-based state initialization for the logical $|\bar{0}\rangle$ state in the Steane code and (b) for the logical $|\bar{+}\rangle$ state in the qRM code. The initialization consist of a non-fault tolerant encoding followed by a verification step, which detects propagated uncorrectable $X$-errors for (a) and $Z$-errors for (b), respectively.}
    \label{fig:merge_init}
\end{figure*}

The stabilizer group $S_{\text{2D}}$ of the Steane [[7,1,3]] code is generated by
\begin{equation}
\begin{split}
    p_1^Z&=Z_1Z_2Z_6Z_7 \quad , \quad p_1^X=X_1X_2X_6X_7 \quad,\\
    p_2^Z&=Z_2Z_3Z_4Z_7 \quad , \quad p_2^X=X_2X_3X_4X_7\quad  ,\\
    p_3^Z&=Z_4Z_5Z_6Z_7 \quad , \quad p_3^X=X_4X_5X_6X_7 \quad.
\label{eq:steane_plaquettes}
\end{split}
\end{equation}
These stabilizer generators correspond to weight-$4$ $2$-cells or plaquettes, as it can be observed from Fig.~\ref{fig:Steane}. Note that since the Steane code is a self-dual code and the stabilizer weights are $4$, the $Y$-type stabilizers are $p_i^Y=p_i^Xp_i^Z$ for $i=1,2,3$. Minimum weight representatives of the Pauli logical operators $\bar{X}$, $\bar{Y}$ and $\bar{Z}$ are
\begin{equation}
    \bar{X}=X_1X_2X_3 \quad , \quad \bar{Y}=-Y_1Y_2Y_3 \quad , \quad \bar{Z}=Z_1Z_2Z_3 \:.
\label{eq:Steane_logicals}
\end{equation}
The logical $|\bar{0}\rangle$ state is thus a $+1$ eigenstate of $\bar{Z}|\bar{0}\rangle=|\bar{0}\rangle$ as well as of the stabilizer group generators. In the protocol of Fig.~\ref{fig:logical_circuit}, the $|\bar{0}\rangle$ is initialized using the technique of~\cite{goto2016minimizing}, which is shown in Fig.~\ref{fig:merge_init}a.

\begin{figure}[!ht]
    \centering
    \includegraphics[width=0.3\textwidth]{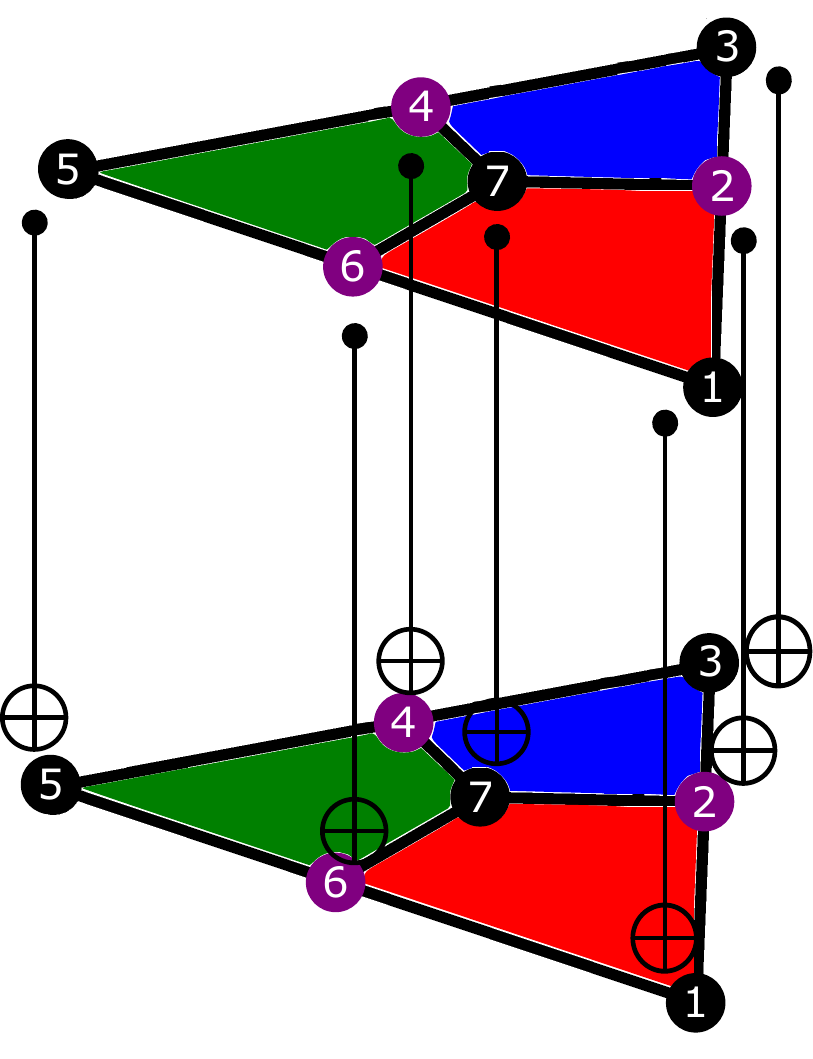}
    \caption{Logical $\overline{\text{CNOT}_{s}}$ gate between Steane codes, consisting of physical CNOT gates between corresponding physical qubits.}
    \label{fig:cnot_Steane}
\end{figure}

The logical $\overline{\text{CNOT}_{s}}$ gate utilized in the two-copy experiment of Fig.~\ref{fig:two_copy_protocol}b is shown in Fig.~\ref{fig:cnot_Steane}.

\subsection{Quantum Reed-Muller code}
\label{app:qRM_code}
The stabilizer group $S_{\text{3D}}$ of the quantum Reed-Muller [[15,1,3]] (qRM) code is generated by $Z$-type generators
\begin{equation}
\begin{split}
    p_1&=Z_1Z_2Z_6Z_7 \quad \:\:\:\:, \quad p_{10}=Z_6Z_7Z_{13}Z_{14}  \quad \:\:\:\: ,\\
    p_2&=Z_2Z_3Z_4Z_7 \quad \:\:\:\:, \quad p_{11}=Z_2Z_7Z_9Z_{14}  \quad \:\:\:\:\:\:,\\
    p_3&=Z_4Z_5Z_6Z_7 \quad \:\:\:\:, \quad p_{12}=Z_4Z_7Z_{11}Z_{14} \quad \:\:\:\:,\\
    p_4&=Z_1Z_6Z_8Z_{13} \quad \:\:, \quad p_{13}=Z_8Z_{12}Z_{13}Z_{15}\quad \:\:,\\
    p_5&=Z_1Z_2Z_8Z_9 \quad \:\:\:\:, \quad p_{14}=Z_8Z_9Z_{10}Z_{15}\quad \:\:\:\:,\\
    p_6&=Z_2Z_3Z_9Z_{10} \quad \:\:, \quad p_{15}=Z_{10}Z_{11}Z_{12}Z_{15}\quad ,\\
    p_7&=Z_3Z_4Z_{10}Z_{11} \quad , \quad p_{16}=Z_8Z_9Z_{13}Z_{14}\quad \:\:\:\:,\\
    p_8&=Z_4Z_5Z_{11}Z_{12} \quad , \quad p_{17}=Z_9Z_{10}Z_{11}Z_{14}\quad \:\:,\\
    p_9&=Z_5Z_6Z_{12}Z_{13} \quad , \quad p_{18}=Z_{11}Z_{12}Z_{13}Z_{14} \quad ,
\label{eq:plautettes_qRM}
\end{split}
\end{equation}
as well as by $X$-type generators,
\begin{equation}
    \begin{split}
        c_1&=X_1X_2X_6X_7X_8X_9X_{13}X_{14}\quad \:\:\:\:\:\:\:,\\
        c_2&=X_4X_5X_6X_7X_{11}X_{12}X_{13}X_{14}\quad \:\:\: ,\\
        c_3&=X_2X_3X_4X_7X_9X_{10}X_{11}X_{14}\quad \:\:\:\:\:,\\
        c_4&=X_8X_9X_{10}X_{11}X_{12}X_{13}X_{14}X_{15}\quad .
    \label{eq:3cells}
    \end{split}
\end{equation}
From the total of $18$ weight-$4$ plaquette $Z$-type stabilizer generators, only $10$ of them are independent, for instance $\{p_1,p_2,p_3,p_7,p_8,p_9,p_{13},p_{16},p_{17},p_{18}\}$~\cite{Daguerre:2024gjd}. The $X$-type generators correspond to weight-$8$ $3$-cells, as it can be observed from Fig.~\ref{fig:qRM}.  Minimum weight representatives of the Pauli logical operators $\bar{X}$ and $\bar{Z}$ are
\begin{equation}
    \bar{X}=X_1X_2X_3X_4X_5X_6X_7 \quad , \quad \bar{Z}=Z_1Z_2Z_3 \:.
\end{equation}
The logical $|\bar{+}\rangle$ state is thus a $+1$ eigenstate of $\bar{X}|\bar{+}\rangle=|\bar{+}\rangle$ as well as of the stabilizer group generators. In the protocol of Fig.~\ref{fig:logical_circuit}, the $|\bar{+}\rangle$ is initialized using the technique of~\cite{Butt2024}, which is shown in Fig.~\ref{fig:merge_init}b.

\begin{figure*}[!ht]
    \centering
    \includegraphics[width=1\textwidth]{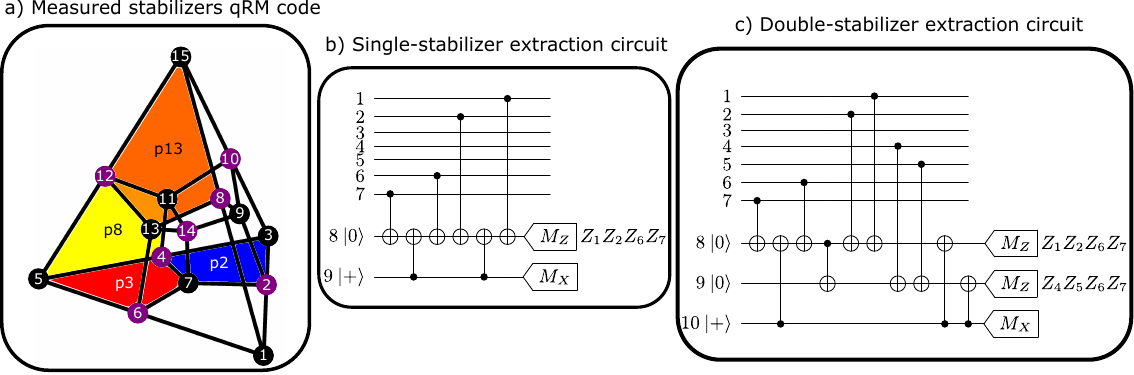}
    \caption{(a) Set of $Z$-type plaquette stabilizers of the qRM code $\{p_2,p_3,p_8,p_{13}\}$, which can detect high-weight uncorrectable $X$-errors originating from the encoding of the logical $|\bar{+}\rangle$ state, Fig.~\ref{fig:merge_init}b. (b) Flag-based single-stabilizer extraction circuit, used for measuring the $Z$-type plaquettes $p_8$ and $p_{13}$. (3) Parallel-flag-based double-stabilizer extraction circuit, used for simultaneously measuring the $Z$-type plaquettes $\{p_2,p_3\}$.}
    \label{fig:stab_qRM}
\end{figure*}

Since uncorrectable higher-weight $X$-type errors are not detected in the initialization of the $|\bar{+}\rangle$ of Fig.~\ref{fig:merge_init}b, they can lead to a logical failure. Hence, a set of independent $Z$-type stabilizers should be measured before applying the logical $\bar{T}$ gates in the qRM code. However, if we are only interested in detecting uncorrectable weight-$2$ or higher $X$-type errors, it turns out that $4$ stabilizers is enough~\cite{Daguerre:2024gjd}. Indeed, we measure the following stabilizers before applying the logical $\bar{T}$ gate $\{p_2,p_3,p_8,p_{13}\}$, see Fig.~\ref{fig:stab_qRM}a. The plaquettes $p_{13}$ and $p_8$ are measured first, utilizing the flag-based stabilizer extraction circuit of~\cite{Chao2018} (see Fig.~\ref{fig:stab_qRM}b), and finally $p_2$ and $p_3$ are simultaneously measured using the parallel flag-based stabilizer extraction circuit of ~\cite{Chao2018}, see Fig.~\ref{fig:stab_qRM}c.


\section{Logical state certification method}
\label{app:tomography}
In this Appendix we prove a lower bound for the fidelity $F(\bar{\rho},\bar{\sigma}_T)=\text{Tr}(\bar{\rho}\bar{\sigma}_T)$, between an experimental state $\bar{\rho}$ and a target ideal pure state $\bar{\sigma}_T$, both with Bloch sphere decompositions
\begin{equation}
    \bar{\rho}=\frac{I+\vec{v}\cdot \vec{\sigma}}{2} \quad,\quad \bar{\sigma}_T=\frac{I+\vec{u}\cdot \vec{\sigma}}{2}\quad, \quad |\vec{u}|^2=1\:.
\end{equation}
We define the difference between their vector representation $\vec{\delta}$ as
\begin{equation}
    \vec{v}=\vec{u}+\vec{\delta}\:.
\end{equation}
The vector $\vec{v}=\big(\langle \bar{X}\rangle_{\bar{\rho}},\langle \bar{Y}\rangle_{\bar{\rho}},\langle \bar{Z}\rangle_{\bar{\rho}}\big)$. For a magic $|\bar{T}\rangle \langle \bar{T}|$ state $\vec{u}=\big(\frac{1}{\sqrt{2}},\frac{1}{\sqrt{2}},0\big)$. The fidelity $F(\bar{\rho},\bar{\sigma}_T)$ is
\begin{equation}
    F(\bar{\rho},\bar{\sigma}_T)=\frac{1}{2}\left(1+\vec{u}\cdot \vec{v}\right)=1+\frac{1}{2}\vec{u}\cdot \vec{\delta}\quad.
\label{eq:fid_delta}
\end{equation}
On the other hand, the trace of the state squared is
\begin{equation}
    \text{Tr}(\bar{\rho}^2)=\frac{1}{2}(1+|\vec{v}|^2)=1+\frac{1}{2}|\vec{\delta}|^2+\vec{u}\cdot \vec{\delta}\quad.
\label{eq:rho_quad}
\end{equation}
Thus, combining (\ref{eq:fid_delta}) and (\ref{eq:rho_quad}) yields
\begin{equation}
    \text{Tr}(\bar{\rho}^2)=1+\frac{1}{2}|\vec{\delta}|^2+2\Big(F(\bar{\rho},\bar{\sigma}_T)-1\Big)\:.
    \label{eq:trace_fid}
\end{equation}

The values of $\vec{\delta}$ can be obtained from the single-copy experiment. However, the trace of the state squared has to be estimated from an experiment involving at least two-copies, see Section~\ref{sec:state_tom}. Indeed, similarly to the SWAP test~\cite{barenco1997stabilization} , which determines the purity of a state, the projection onto the singlet (anti-symmetric) state $|\overline{\Phi}_1 \rangle =\frac{1}{\sqrt{2}}(|\overline{01}\rangle-|\overline{10}\rangle)$ is equal to
\begin{equation}
\begin{split}
    \text{Tr}(\bar{\rho}_1\bar{\rho}_2)&=\text{Tr}(\overline{\text{SWAP}}(\bar{\rho}_1 \otimes \bar{\rho}_2))\\
    &=1-2\langle \overline{\Phi}_1 | \bar{\rho}_1 \otimes \bar{\rho}_2|\overline{\Phi}_1\rangle\:.
\end{split}
\end{equation}
Note that we do not assume that both copies $\bar{\rho}_1$ and $\bar{\rho}_2$ are identical, even though they were produced using the same quantum circuits. The projection $\epsilon$ is defined as
\begin{equation}
\begin{split}
    \epsilon &\equiv \langle \overline{\Phi}_1 | \bar{\rho}_1 \otimes \bar{\rho}_2|\overline{\Phi}_1\rangle\\ &=\langle \overline{-1}|\overline{\text{CNOT}_{s}} (\bar{\rho}_1 \otimes \bar{\rho}_2)\overline{\text{CNOT}_{s}}|\overline{-1}\rangle\:,
\end{split}
\end{equation}
where $|\overline{-1}\rangle \equiv |\overline{-}\rangle_1 |\bar{1}\rangle_2$. Due to the Cauchy-Schwarz inequality
\begin{equation}
    \sqrt{\text{Tr}(\bar{\rho}_1^2)\text{Tr}(\bar{\rho}_2^2)}\geq \text{Tr}(\bar{\rho}_1\bar{\rho}_2)= 1-2\epsilon \:.
\end{equation}
Moreover, because of the AM-GM inequality
\begin{equation}
    \frac{\text{Tr}(\bar{\rho}_1^2)+\text{Tr}(\bar{\rho}_2^2)}{2}\geq \sqrt{\text{Tr}(\bar{\rho}_1^2)\text{Tr}(\bar{\rho}_2^2)}\geq 1-2\epsilon.
\end{equation}
Then, $1-2\epsilon$ is a lower bound for the maximum (or average) trace of the experimental state squared
\begin{equation}
    \text{max}\Big\{\text{Tr}(\bar{\rho}_i^2)\Big\}\geq 1-2\epsilon\:.
\label{eq:bound_trace}
\end{equation}
Finally, combining (\ref{eq:trace_fid}) and (\ref{eq:bound_trace}) we prove the lower bound on the fidelity for at least one copy 
\begin{equation}
     F(\bar{\rho},\bar{\sigma}_T) \geq  1-\left(\epsilon+\frac{\delta_X^2+\delta_Y^2+\delta_Z^2}{4}\right)\:.
\end{equation}


\end{document}